\documentstyle[12pt]{article}

\begin{document}

\title{Anholonomic Frames and Thermodynamic Geometry
 of 3D  Black Holes}
\author{Sergiu I. Vacaru\thanks{%
e--mail:\ vacaru@fisica.ist.utl.pt,
sergiu$_{-}$vacaru@yahoo.com}\ , Panyiotis Stavrinos \thanks{%
e--mail:\ pstavrin@cc.uoa.gr}\quad and Denis Gontsa
\thanks{e--mail:\ d$_{-}$gontsa@yahoo.com } \quad \\
\\
{\small * \ Physics Department, CSU Fresno,\ Fresno, CA 93740-8031, USA, }\\
{\small \& }\\
{\small Centro Multidisciplinar de Astrofisica - CENTRA,
Departamento de  Fisica,}\\
{\small Instituto Superior Tecnico, Av. Rovisco Pais 1, Lisboa,
1049-001,
Portugal,}\\
{\small {---}} \\
{\small $\dag$  Department of Mathematics, University of Athens,}\\
{\small 15784 Panepistimiopolis, Athens, Greece} \\
and \\
{\small $\ddag$ \ Department of Physics, St. Petersbourg State University }\\
{\small P. O. Box 122, Petergoff, St. Petersbourg, 198904,
Russia  }}
\date{May 31, 2002}
\maketitle

\begin{abstract}
 We study new classes three dimensional  black hole solutions of Einstein
 equations written in two holonomic and one anholonomic variables with respect to
  anholonomic frames.
 Thermodynamic properties of such $(2+1)$--black holes with
 generic local  anisotropy (having elliptic horizons)
  are studied by applying geometric methods.
 The corresponding thermodynamic systems are
 three dimensional with entropy $S$ being a hypersurface function on
 mass $M,$ anisotropy angle $\theta$  and eccentricity of elliptic
 deformations $\varepsilon.$ Two--dimensional curved thermodynamic
 geometries for locally anistropic deformed black holes are constructed
 after integration on anisotropic parameter $\theta$. Two approaches,
 the first one  based on two--dimensional hypersurface parametric
 geometry and the second one developed in a Ruppeiner--Mrugala--Janyszek
 fashion, are analyzed.  The thermodynamic  curvatures are computed and
 the critical points of curvature vanishing are defined.
 \end{abstract}


\section{Introduction}

This is the second paper in a series in which we examine black holes for
spacetimes with generic local anisotropy. Such spacetimes are usual
pseudo--Riemannian spaces for which an anholonomic frame structure  By using
moving anholonomic frames one can construct solutions of Einstein equations
with deformed spherical  symmetries (for instance, black holes with elliptic
horizons  (in three dimensions, 3D), black tora and another type
configurations) which are locally anisotropic \cite{v3,v6}.

In the first paper \cite{v4} (hereafter referred to as Paper I)
we analyzed the low--dimension\-al locally anisotropic gravity (we
shall use terms like local\-ly an\-isotropic gravity, locally
an\-isotropic spacetime, locally anisotropic geometry, locally
an\-isotropic black holes and so on) and constructed new classes
of locally anisotropic $(2+1)$--dimensional black hole solutions.
We emphasize that in this work the splitting $(2+1)$  points not
to a space--time decomposition, but to a spacetime distribution in
two isotropic and one anisotropic coordinate.

In particular, it was shown following \cite{v2} how black holes can recast
in a new fashion in generalized Kaluza--Klein spaces and emphasized that
such type solutions can be considered in the framework of usual Einstein
gravity on anholonomic manifolds. We discussed the physical properties of $%
(2+1)$--dimensional black holes with locally anisotropic matter, induced by
a rotating null fluid and by an inhomogeneous and non--static collapsing
null fluid, and examined the vacuum polarization of locally anisotropic
spacetime by non--rotating black holes with ellipsoidal horizon and by
rotating locally anisotropic black holes with time oscillating and
ellipsoidal horizons. It was concluded that a general approach to the
locally anisotropic black holes should be based on a kind of nonequilibrium
thermodynamics of such objects imbedded into locally anisotropic spacetime
background. Nevertheless, we proved that for the simplest type of locally
anisotropic black holes theirs thermodynamics could be defined in the
neighborhoods of some equilibrium states when the horizons are deformed but
constant with respect to a frame base locally adapted to a nonlinear
connection structure which model a locally anisotropic configuration.

In this paper we will specialize to the geometric thermodynamics of, for
simplicity non--rotating, locally anisotropic black holes with elliptical
horizons. We follow the notations and results from the Paper I which are
reestablished in a manner compatible in the locally isotropic thermodynamic
\cite{cai} and spacetime \cite{btz} limits with the
Banados--Teitelboim--Zanelli (BTZ) black hole. This new approach (to black
hole physics) is possible for locally anisotropic spacetimes and is based on
classical results \cite{gp,klim,kre,sien}.

Since the seminal works of Bekenstein \cite{bek}, Bardeen, Carter and
Hawking \cite{bch} and Hawking \cite{haw1}, black holes were shown to have
properties very similar to those of ordinary thermodynamics. One was treated
the surface gravity on the event horizon as the temperature of the black
hole and proved that a quarter of the event horizon area corresponds to the
entropy of black holes. At present time it is widely believed that a black
hole is a thermodynamic system (in spite of the fact that one have been
developed a number of realizations of thermodynamics involving radiation)
and the problem of statistical interpretation of the black hole entropy is
one of the most fascinating subjects of modern investigations in
gravitational and string theories.

In parallel to the 'thermodynamilazation' of black hole physics
one have developed a new approach to the classical thermodynamics
based of Riemannian geometry and its generalizations (a review on
this subject is contained in Ref. \cite{rup}). Here is to be
emphasized that geometrical methods have always played an
important role in thermodynamics (see, for instance, a work by
Blaschke \cite{bla} from 1923). Buchdahl used in 1966 a Euclidean
metric in thermodynamics \cite{buch} and then Weinhold considered
a sort of Riemannian metric \cite{wein}. It is considered that
the Weinhold's metric has not physical interpretation in the
context of purely equilibrium thermodynamics \cite{rup0,rup} and
Ruppeiner introduced a new metric (related via the temperature
$T$ as the conformal factor with the Weinhold's metric).

The thermodynamical geometry was generalized in various directions, for
instance, by Janyszek and Mrugala \cite{jm1,jm2,m3} even to discussions of
applications of Finsler geometry in thermodynamic fluctuation theory and for
nonequilibrium thermodynamics \cite{sien}.

Our goal will be to provide a characterization of thermodynamics of $(2+1)$%
--dimensional locally anisotropic black holes with elliptical (constant in
time) horizon obtained in \cite{v3,v4}. From one point of view we shall
consider the thermodynamic space of such objects (locally anisotropic black
holes in local equilibrium with locally anisotropic spacetime ether) to
depend on parameter of anisotropy, the angle $\theta ,$ and on deformation
parameter, the eccentricity $\varepsilon .$ From another point, after we
shall integrate the formulas on $\theta ,$ the thermodynamic geometry will
be considered in a usual two--dimensional Ruppeiner--Mrugala--Janyszek
fashion. The main result of this work are the computation of thermodynamic
curvatures and the proof that constant in time elliptic locally anisotropic
black holes have critical points of vanishing of curvatures (under both
approaches to two--dimensional thermodynamic geometry) for some values of
eccentricity, i. e. for under corresponding deformations of locally
anisotropic spacetimes.

The paper is organized as follows. In Sec. 2, we briefly review
the geometry pseudo--Riemannian spaces provided with anholonimic
frame and
associated nonlinear connection structure and present the $(2+1)$%
--dimensional constant in time elliptic black hole solution. In
Sec. 3, we state the thermodynamics of nearly equilibrium
stationary locally anisotropic black holes and establish the
basic thermodynamic law and relations. In Sec. 4 we develop two
approaches to the thermodynamic geometry of locally anisotropic
black holes, compute thermodynamic curvatures and the equations
for critical points of vanishing of curvatures for some values of
eccentricity. In Sec. 5, we draw a discussion and conclusions.

\section{Locally Anisotropic Spacetimes and Black Holes}

In this section we outline for further applications the basic results on $%
(2+1)$--dimensional locally anisotropic spacetimes and locally anisotropic
black hole solutions \cite{v3,v4}.

\subsection{Anholonomic frames and nonlinear connections in $(2+1)$%
--dimensional spacetimes}

A (2+1)--dimensional locally anisotropic spacetime is defined as a 3D
pseudo--Riemannian space provided with a structure of anholonomic frame with
two holonomic coordinates $x^i,i=1,2$ and one anholonomic coordinate $y,$
for which $u=(x,y)=\{u^\alpha =(x^i,y)\},$ the Greek indices run values $%
\alpha =1,2,3,$ when $u^3=y.$ We shall use also underlined indices, for
instance $\underline{\alpha },\underline{i},$ in order to emphasize that
some tensors are given with respect to a local coordinate base $\partial _{%
\underline{\alpha }}=\partial /\partial u^{\underline{\alpha }}.$

An anholonomic frame structure of triads (dreibein) is given by a set of
three independent basis fields
\[
e_\alpha (u)=e_\alpha ^{\underline{\alpha }}(u)\partial _{\underline{\alpha }%
}
\]
which satisfy the relations
\[
e_\alpha e_\beta -e_\beta e_\alpha =w_{\ \alpha \beta }^\gamma e_\gamma ,
\]
where $w_{\ \alpha \beta }^\gamma =w_{\ \alpha \beta }^\gamma
(u)$ are called anholonomy coefficients.

We investigate anholonomic structures with mixed holonomic and
anholonomic tri\-ads when
\[
e_\alpha ^{\underline{\alpha }}(u)=\{e_j^{\underline{i}}=\delta _j^i,e_j^{%
\underline{3}}=N_j^3(u)=w_i(u),e_3^{\underline{3}}=1\}.
\]
In this case we have to apply 'elongated' by N--coefficients operators
instead of usual local coordinate basis $\partial _\alpha =\partial
/\partial u^\alpha $ and $d^\alpha =du^\alpha ,$ (for simplicity we shall
omit underling of indices if this does not result in ambiguities):
\begin{eqnarray}
\delta _\alpha & = & (\delta _i,\partial _{(y)})=\frac \delta {\partial
u^\alpha }   \label{2.1} \\
& = & \left( \frac \delta {\partial x^i}=\frac \partial {\partial x^i}-
w_i\left( x^j,y\right) \frac \partial {\partial y}, \partial _{(y)} = \frac %
\partial {\partial y}\right)  \nonumber
\end{eqnarray}
and their duals
\begin{eqnarray}
\delta ^\beta & = & \left( d^i,\delta ^{(y)}\right) = \delta u^\beta
\label{2.2} \\
& = & \left( d^i = dx^i, \delta ^{(y)} = \delta y=dy+w_k\left( x^j,y\right)
dx^k\right) .  \nonumber
\end{eqnarray}

The coefficients $N=\{N_i^3\left( x,y\right) =w_i\left(
x^j,y\right) \},$ are associated to a nonlinear connection (in
brief, N--connection, see \cite {barthel}) structure which on
pseudo-Riemannian spaces defines a locally anisotropic, or
equivalently, mixed holo\-no\-mic--anholonomic structure. The
geometry of N--connection was investigated for vector bundles and
generalized Finsler geometry \cite{ma} and for superspaces and
locally anisotropic (super)gravity and string theory \cite{v2}
with applications in general relativity, extra dimension gravity
and formulation of locally anisotropic kinetics and
thermodynamics on curved spacetimes \cite{v3,v4,v6}. In this paper
(following the Paper I) we restrict our considerations to the
symplest case with one anholonomic (anisotropic) coordinate when
the N--connection is associated to a subclass of anholonomic
triads (\ref{2.1}), and/or (\ref{2.2}), defining some locally
anisotropic frames (in brief, anholonomic basis, anholonomic
frames).

With respect to a fixed structure of locally anisotropic bases and their
tensor products we can construct distinguished, by N--connection, tensor
algebras and various geometric objects (in brief, one writes d--tensors,
d--metrics, d--connections and so on).

A symmetrical locally anisotropic metric, or d--metric, could be written
with respect to an anhlonomic basis (\ref{2.2}) 
as
\begin{eqnarray}
\delta s^2 &= & g_{\alpha \beta }\left( u^\tau \right) \delta u^\alpha
\delta u^\beta   \label{2.3} \\
& = & g_{ij}\left( x^k,y\right) dx^idx^j+h\left( x^k,y\right) \left( \delta
y\right) ^2.  \nonumber
\end{eqnarray}
We note that the anisotropic coordinate $y$ could be both type time--like $%
(y=t,$ or space--like coordinate, for instance, $y=r,$ radial coordinate, or
$y=\theta ,$ angular coordinate).

\subsection{Non--rotating black holes with ellipsoidal horizon}

Let us consider a 3D locally anisotropic spacetime provided with local space
coordinates $x^{1}=r,\ x^{2}=\theta $ when as the anisotropic direction is
chosen the time like coordinate, $y=t.$ We proved (see the Paper I) that a
d--metric of type (\ref{2.3}), 
\begin{equation}
\delta s^{2}=\Omega ^{2}\left( r,\theta \right) \left[ a(r)dr^{2}+b\left(
r,\theta \right) d\theta ^{2}+h(r,\theta )\delta t^{2}\right] ,  \label{2.4}
\end{equation}
where
\begin{eqnarray*}
\delta t &=&dt+w_{1}\left( r,\theta \right) dr+w_{2}(r,\theta )d\theta , \\
w_{1} &=&\partial _{r}\ln \left| \ln \Omega \right| ,w_{2}=\partial _{\theta
}\ln \left| \ln \Omega \right| ,
\end{eqnarray*}
for $\Omega ^{2}=\pm h(r,\theta ),$ satisfies the system of vacuum locally
anisotropic gravitational equations with cosmolodgical constant $\Lambda
_{\lbrack 0]},$%
\[
R_{\alpha \beta }-\frac{1}{2}g_{\alpha \beta }R-\Lambda _{\lbrack
0]}g_{\alpha \beta }=0
\]
if
\[
a\left( r\right) =4r^{2}|\Lambda _{0}|,b(r,\theta )=\frac{4}{|\Lambda _{0}|}%
\Lambda ^{2}(\theta )\left[ r_{+}^{2}(\theta )-r^{2}\right] ^{2}
\]
and
\begin{equation}
h\left( r,\theta \right) =-\frac{4}{|\Lambda _{0}|r^{2}}\Lambda ^{3}(\theta )%
\left[ r_{+}^{2}(\theta )-r^{2}\right] ^{3}.  \label{2.5}
\end{equation}
The functions $a(r),b\left( r,\theta \right) $ and $h(x^{i},y)$
and the coefficients of nonlinear connection\\ $w_{i}(r,\theta
,t)$ (for this class of solutions being arbitrary prescribed
functions) were defined as to have compatibility with the locally
isotropic limit.

We construct a black hole like solution with elliptical horizon $%
r^2=r_{+}^2(\theta ),$ on which the function (\ref{2.5}) vanishes  if we
chose
\begin{equation}  \label{2.7}
r_{+}^2(\theta )=\frac{p^2}{\left[ 1+\varepsilon \cos \theta \right] ^2}.
\end{equation}
where $p$ is the ellipse parameter and $\dot \varepsilon $ is the
eccentricity. We have\ to identify
\[
p^2=r_{+[0]}^2=-M_0/\Lambda _0,
\]
where\ $r_{+[0]},M_0$ and $\Lambda _0$ are respectively the horizon radius,
mass parameter and the cosmological constant of the non--rotating BTZ
solution \cite{btz} if we wont to have a connection with locally isotropic
limit with $\varepsilon \rightarrow 0.$ In the simplest case we can consider
that the elliptic horizon (\ref{2.7}) is modeled by an anisotropic mass
\begin{equation}  \label{3.1}
M\left( \theta ,\varepsilon \right) =\frac{M_0}{2\pi \left( 1+\varepsilon
\cos \left( \theta -\theta _0\right) \right) ^2}=\frac{r_{+}^2}{2\pi }
\end{equation}
and constant effective cosmological constant, $\Lambda (\theta )\simeq
\Lambda _0.$ The coefficient $2\pi \,$ was introduced in order to have the
limit
\begin{equation}  \label{3.2}
\lim _{\varepsilon \rightarrow 0}2\int\limits_0^\pi M\left( \theta
,\varepsilon \right) d\theta =M_0.
\end{equation}
Throughout this paper, the units $c=\hbar =k_B=1$ will be used, but we shall
consider that for an locally anisotropically renormalized gravitational
constant $8G_{(gr)}^{(a)}\neq 1,$ see \cite{v4}.

\section{On the Thermodynamics of Elliptical Black Holes}

In this paper we will be interested in thermodynamics of locally anisotropic
black holes defined by a d--metric (\ref{2.4}). 

The Hawking temperature $T\left( \theta ,\varepsilon \right) $ of a locally
an\-iso\-trop\-ic black hole is anisotropic and is computed by using the
anisotropic mass 
(\ref{3.1}):
\begin{equation}  \label{3.3}
T\left( \theta ,\varepsilon \right) =\frac{M\left( \theta ,\varepsilon
\right) }{2\pi r_{+}\left( \theta ,\varepsilon \right) }=\frac{r_{+}\left(
\theta ,\varepsilon \right) }{4\pi ^2}>0.
\end{equation}

The two parametric analog of the Bekenstein--Hawking entropy is to be
defined as
\begin{equation}  \label{3.4}
S\left( \theta ,\varepsilon \right) =4\pi r_{+}=\sqrt{32\pi ^3}\sqrt{M\left(
\theta ,\varepsilon \right) }
\end{equation}

The introduced thermodynamic quantities obey the first law of thermodynamics
(under the supposition that the system is in local equilibrium under the
variation of parameters $\left( \theta ,\varepsilon \right) $)
\begin{equation}  \label{3.5}
\Delta M\left( \theta ,\varepsilon \right) =T\left( \theta ,\varepsilon
\right) \Delta S,
\end{equation}
where the variation of entropy is
\[
\Delta S=4\pi \Delta r_{+}=4\pi \frac{1 }{\sqrt{M\left( \theta ,\varepsilon
\right) }}\left( \frac{\partial M}{\partial \theta }\Delta \theta +\frac{%
\partial M}{\partial \varepsilon }\Delta \varepsilon \right) .
\]
According to the formula $C=\left( \partial m/\partial T\right) $ we can
compute the heat capacity
\[
C=2\pi r_{+}\left( \theta ,\varepsilon \right) =2\pi \sqrt{M\left( \theta
,\varepsilon \right) }.
\]
Because of $C>0$ always holds the temeperature is increasing with the mass.

The formulas (\ref{3.1})--(\ref{3.5}) 
can be integrated on angular variable $\theta $ in order to obtain some
thermodynamic relations for black holes with elliptic horizon depending only
on deformation parameter, the eccentricity $\varepsilon .$

For a elliptically deformed black hole with the outer horizon $r_{+}$ given
by formula (\ref{3.4}) 
the depending on eccentricity\cite{v4} Bekenstein--Hawking entropy is
computed as
\[
S^{(a)}\left( \varepsilon \right) =\frac{L_{+}}{4G_{(gr)}^{(a)}},
\]
were
\[
L_{+}\left( \varepsilon \right) =4\int\limits_0^{\pi /2}r_{+}\left( \theta
,\varepsilon \right) d\theta
\]
is the length of ellipse's perimeter and $G_{(gr)}^{(a)}$ is the three
dimensional gravitational coupling constant in locally anisotropic media
(the index $\left( a\right) $ points to locally anisotropic
renormalizations), and has the value
\begin{equation}  \label{3.6}
S^{(a)}\left( \varepsilon \right) =\frac{2p}{G_{(gr)}^{(a)}\sqrt{%
1-\varepsilon ^2}}arctg\sqrt{\frac{1-\varepsilon }{1+\varepsilon }}.
\end{equation}
If the eccentricity vanishes, $\varepsilon =0,$ we obtain the locally
isotropic formula with $p$ being the radius of the horizon circumference,
but the constant $G_{(gr)}^{(a)}$ could be locally anisotropic renormalized.

The total mass of a locally anisotropic black hole of eccentricity $%
\varepsilon $ is found by integrating (\ref{3.1}) 
on angle $\theta :$%
\begin{equation}  \label{3.7}
M\left( \varepsilon \right) =\frac{M_0}{\left( 1-\varepsilon ^2\right) ^{3/2}%
}
\end{equation}
which satisfies the condition (\ref{3.2}). 

The integrated on angular variable $\theta $ temperature $T\left(
\varepsilon \right) $ is to by defined by using $T\left( \theta ,\varepsilon
\right) $ from (\ref{3.3}), 
\begin{equation}  \label{3.8}
T\left( \varepsilon \right) =4\int\limits_0^{\pi /2}T\left( \theta
,\varepsilon \right) d\theta =\frac{2\sqrt{M_0}}{\pi ^2\sqrt{1-\varepsilon ^2%
}}arctg\sqrt{\frac{1-\varepsilon }{1+\varepsilon }.}
\end{equation}

Formulas (\ref{3.6})--(\ref{3.8}) 
describes the thermodynamics of $\varepsilon $--deformed black holes.

Finally, in this section, we note that a black hole with elliptic horizon is
to be considered as a thermodynamic subsystem placed into the anisotropic
ether bath of spacetime. To the locally anisotropic ether one associates a
continuous locally anisotropic medium assumed to be in local equilibrium.
The locally anisotropic black hole subsystem is considered as a subsystem
described by thermodynamic variables which are continuous field on variables
$\left( \theta ,\varepsilon \right) ,$ or in the simplest case when one have
integrated on $\theta ,$ on $\varepsilon .$ It will be our first task to
establish some parametric thermodynamic relations between the mass $m\left(
\theta ,\varepsilon \right) $ (equivalently, the internal locally
anisotropic black hole energy), temperature $T\left( \theta ,\varepsilon
\right) $ and entropy $S\left( \theta ,\varepsilon \right) .$

\section{Thermodynamic Metrics and Curvatures of An\-iso\-tro\-pic Black Holes}

We emphasize in this paper two approaches to the thermodynamic geometry of
nearly equilibrium locally anisotropic black holes based on their
thermodynamics. The first one is to consider the thermodynamic space as
depending locally on two parameters $\theta$ and $\varepsilon$ and to
compute the corresponding metric and curvature following standard formulas
from curved bidimensional hypersurface Riemannian geometry. The second
possibility is to take as basic the Ruppeiner metric in the thermodynamic
space with coordinates $(M, \varepsilon ),$ in a manner proposed in Ref.
\cite{cai} with that difference that as the extensive coordinate is taken
the black hole eccentricity $\varepsilon$ (instead of the usual angular
momentum $J$ for isotropic $(2+1)$--black holes). Of course, in this case we
shall background our thermodynamic geometric constructions starting from the
relations 
(\ref{3.6})--(\ref{3.8}).

\subsection{The thermodynamic parametric geometry}

Let us consider the thermodynamic parametric geometry of the elliptic
(2+1)--dimensional black hole based on its thermodynamics given by formulas (%
\ref{3.1})--(\ref{3.5}). 

Rewriting equations (\ref{3.5}, 
we have
\[
\Delta S=\beta \left( \theta ,\varepsilon \right) \Delta M\left( \theta
,\varepsilon \right) ,
\]
where $\beta \left( \theta ,\varepsilon \right) =1/T\left( \theta
,\varepsilon \right) $ is the inverse to temperature (\ref{3.3}). 
This case is quite different from that from \cite{cai,ferrara} where there
are considered, respectively, BTZ and dilaton black holes (by introducing
Ruppeiner and Weinhold thermodynamic metrics). Our thermodynamic space is
defined by a hypersurface given by parametric dependencies of mass and
entropy. Having chosen as basic the relative entropy function,
\[
\varsigma =\frac{S\left( \theta ,\varepsilon \right) }{4\pi \sqrt{M_0}}=%
\frac 1{1+\varepsilon \cos \theta },
\]
in the vicinity of a point $P=(0,0),$ when, for simplicity, $\theta _0=0,$
our hypersurface is given locally by conditions
\[
\varsigma =\varsigma \left( \theta ,\varepsilon \right) \mbox{ and }
grad_{|P}\varsigma =0.
\]
For the components of bidimensional metric on the hypersurface we have
\begin{eqnarray}
g_{11} & = & 1+\left( \frac{\partial \varsigma } {\partial \theta }\right)
^2, \ g_{12} = \left( \frac{\partial \varsigma }{\partial \theta }\right)
\left(\frac{\partial \varsigma }{\partial \varepsilon }\right) ,  \nonumber
\\
g_{22} & = & 1+ \left( \frac{\partial \varsigma }{\partial \varsigma }%
\right) ^2,  \nonumber
\end{eqnarray}
The nonvanishing component of curvature tensor in the vicinity of the point $%
P=(0,0)$ is
\[
R_{1212}=\frac{\partial ^2\varsigma }{\partial \theta ^2}\frac{\partial
^2\zeta }{\partial \varepsilon ^2}-\left( \frac{\partial ^2\varsigma }{%
\partial \varepsilon \partial \theta }\right) ^2
\]
and the curvature scalar is
\begin{equation}  \label{4.1}
R=2R_{1212}.
\end{equation}

By straightforward calculations we can find the condition of vanishing of
the curvature (\ref{4.1}) 
when
\begin{equation}  \label{4.2}
\varepsilon _{\pm }=\frac{-1\pm (2-\cos ^2\theta )}{\cos \theta \left(
3-\cos ^2\theta \right) }.
\end{equation}
So, the parametric space is separated in subregions with elliptic
eccentricities $0<\varepsilon _{\pm }<0$ and $\theta $ satisfying conditions
(\ref{4.2}). 

Ruppeiner suggested that the curvature of thermodynamic space is a measure
of the smallest volume where classical thermodynamic theory based on the
assumption of a uniform environment could conceivably work and that near the
critical point it is expected this volume to be proportional to the scalar
curvature \cite{rup}. There were also proposed geometric equations relating
the thermodynamic curvature via inverse relations to free energy. Our
definition of thermodynamic metric and curvature in parametric spaces
differs from that of Ruppeiner or Weinhold and it is obvious that relations
of type (\ref{4.2}) 
(stating the conditions of vanishing of curvature) could be related with
some conditions for stability of thermodynamic space under variations of
eccentricity $\varepsilon $ and anisotropy angle $\theta .$ This
interpretation is very similar to that proposed by Janyszek and Mrugala \cite
{jm1} and supports the viewpoint that the first law of thermodynamics makes
a statement about the first derivatives of the entropy, the second law is
for the second derivatives and the curvature is a statement about the third
derivatives. This treatment holds good also for the parametric thermodynamic
spaces for locally anisotropic black holes.

\subsection{Thermodynamic Metrics and Eccentricity of Black Hole}

A variant of thermodynamic geometry of locally an\-iso\-trop\-ic black holes
could be grounded on integrated on anisotropy angle $\theta $ formulas (\ref
{3.6})--(\ref{3.8}). 
The Ruppeiner metric of elliptic black holes in coordinates $\left(
M,\varepsilon \right) $ is
\begin{equation}  \label{4.3}
ds_R^2=-\left( \frac{\partial ^2S}{\partial M^2}\right) _\varepsilon
dM^2-\left( \frac{\partial ^2S}{\partial \varepsilon ^2}\right)
_Md\varepsilon ^2.
\end{equation}
For our further analysis we shall use dimensionless values $\mu =M\left(
\varepsilon \right) /M_0$ and $\zeta =S^{(a)}G_{gr}^{(a)}/2p$ and consider
instead of (\ref{4.3}) 
the thremodynamic diagonal metrics $g_{ij}\left( a^1,a^2\right) =g_{ij}(\mu
,\varepsilon )$ with components
\begin{equation}  \label{4.4}
g_{11}=-\frac{\partial ^2\zeta }{\partial \mu ^2}=-\zeta _{,11}\mbox{\ and \
}g_{22}=-\frac{\partial ^2\zeta }{\partial \varepsilon ^2}=-\zeta _{,22},
\end{equation}
where by comas we have denoted partial derivatives.

The expressions (\ref{3.6}) 
and (\ref{3.6}) 
are correspondingly rewritten as
\[
\zeta =\frac 1{\sqrt{1-\varepsilon ^2}}arctg\sqrt{\frac{1-\varepsilon }{%
1+\varepsilon }}
\]
and
\[
\mu =\left( 1-\varepsilon ^2\right) ^{-3/2}.
\]

By straightforward calculations we obtain
\begin{eqnarray}
\zeta _{,11} & = & -\frac 19\left( 1-\varepsilon ^2\right) ^{5/2}arctg\sqrt{%
\frac{1-\varepsilon }{1+\varepsilon }}  \nonumber \\
{\ } & + & \frac 1{9\varepsilon }\left( 1-\varepsilon ^2\right) ^3+\frac 1{%
18\varepsilon ^4}\left( 1-\varepsilon ^2\right) ^4  \nonumber
\end{eqnarray}
and
\[
\zeta _{,22}=\frac{1+2\varepsilon ^2}{\left( 1-\varepsilon ^2\right) ^{5/2}}%
arctg\sqrt{\frac{1-\varepsilon }{1+\varepsilon }}-\frac{3\varepsilon }{%
\left( 1-\varepsilon ^2\right) ^2}.
\]

The thermodynamic curvature of metrics of type (\ref{4.4}) 
can be written in terms of second and third derivatives \cite{jm1} by using
third and second order determinants:
\begin{eqnarray}
R & = & \frac 12\left|
\begin{array}{ccc}
-\zeta _{,11} & 0 & -\zeta _{,22} \\
-\zeta _{,111} & -\zeta _{,112} & 0 \\
-\zeta _{,112} & 0 & -\zeta _{,222}
\end{array}
\right| \times \left|
\begin{array}{cc}
-\zeta _{,11} & 0 \\
0 & -\zeta _{,22}
\end{array}
\right| ^{-2}  \nonumber \\
{} & = & -\frac 12\left( \frac 1{\zeta _{,11}}\right) _{,2}\times \left(
\frac{\zeta _{,11}}{\zeta _{,22}}\right) _{,2}.   \label{4.5}
\end{eqnarray}
The conditions of vanishing of thermodynamic curvature (\ref{4.5}) 
are as follows
\begin{equation}  \label{4.6}
\zeta _{,112}\left( \varepsilon _1\right) =0 \mbox{ or } \left( \frac{\zeta
_{,11}}{\zeta _{,22}}\right) _{,2}\left( \varepsilon _2\right) =0
\end{equation}
for some values of eccentricity, $\varepsilon =\varepsilon _1$ or $%
\varepsilon =\varepsilon _2,$ satisfying conditions $0<\varepsilon _1<1$ and
$0<\varepsilon _2<1.$ For small deformations of black holes, i.e. for small
values of eccentricity, we can approximate $\varepsilon _1\approx 1/\sqrt{5.5%
}$ and $\varepsilon _2\approx 1/(18\lambda ),$ where $\lambda $ is a
constant for which $\zeta _{,11}=\lambda \zeta _{,22}$ and the condition $%
0<\varepsilon _2<1$ is satisfied. We omit general formulas for curvature (%
\ref{4.5}) 
and conditions (\ref{4.6}), 
when the critical points $\varepsilon _1$ and/or $\varepsilon _2$ must be
defined from nonlinear equations containing $arctg\sqrt{\frac{1-\varepsilon
}{1+\varepsilon }}$ and powers of $\left( 1-\varepsilon ^2\right) $ and $%
\varepsilon .$

\section{Discussion and Conclusions}

In closing, we would like to discuss the meaning of geometric thermodynamics
following from locally an\-iso\-trop\-ic black holes.

(1) {\it Nonequilibrium thermodynamics of locally\newline
anisotropic black holes in  locally anisotropic spacetimes}. In this paper
and in the Paper I \cite{v4}  we concluded that the thermodynamics in
locally anisotropic spacetimes has a generic nonequilibrium character and
could be developed in a geometric fashion following the approach proposed by
S. Sieniutycz, P. Salamon and R. S. Berry \cite{sien,sal}. This is a new
branch of black hole thermodynamics which should be based on locally
anisotropic nonequilibrium thermodynamics and kinetics \cite{v6}.

(2) {\it Locally Anisotropic Black holes thermodynamics in vicinity of
equilibrium points}. The usual thermodynamical approach in the
Bekenstein--Hawking manner is valid for anisotropic black holes for a
subclass of such physical systems when the hypothesis of local equilibrium
is physically motivated and corresponding renormalizations, by locally
anisotropic spacetime parameters, of thermodynamical values are defined.

(3) {\it The geometric thermodynamics of locally an\-iso\-trop\-ic black
holes with constant in time elliptic horizon} was formulated following two
approaches: for a parametric thermodynamic space depending on anisotropy
angle $\theta$ and eccentricity $\varepsilon$ and in a standard
Ruppeiner--Mrugala--Janyszek fashion, after integration on anisotropy $\theta
$ but maintaining locally anisotropic spacetime deformations on $\varepsilon
.$

(4) {\it The thermodynamic curvatures of locally\newline
anisotropic black holes} were shown to have critical values of eccentricity
when the scalar curvature vanishes. Such type of thermodynamical systems are
rather unusual and a corresponding statistical model is not that for
ordinary systems composed by classical or quantum like gases.

(5) {\it Thermodynamic systems with constraints} require a new
geometric structure in addition to the thermodynamical metrics
which is that of nonlinear connection. We note this object must
be introduced both in spacetime geometry and in thermodynamic
geometry if generic anisotropies and constrained field and/or
thermodynamic behavior are analyzed.

\subsection*{Acknowledgements}

The authors are grateful to  D. Singleton, Radu Miron and M.
Anastasiei for help and collaboration.  The S. V. work is
supported both by a 2000--2001 California State University
Legislative Award and a NATO/Portugal fellowship grant at the
Instituto Superior Tecnico, Lisboa.

\end{document}